\documentclass[print]{mn2e}
\usepackage{epsfig}
\usepackage{color}
\newcommand{\apj}{ApJ}
\newcommand{\aap}{A\&A}
\newcommand{\mn}{MNRAS}
\newcommand{\araa}{ARA\&A}

\newcommand{\pasj}{PASJ}
\newcommand{\nat}{Nature}
\newcommand{\aj}{AJ}

\begin{document}
\title[A past capture event at Sgr A*]{A past capture event at Sagittarius A* inferred from the fluorescent X-ray emission of Sagittarius B clouds}
\author[Yu et al.]{Yun-Wei Yu$^{1,2}$,
K. S. Cheng$^1$, D. O.
Chernyshov$^3$, and V. A. Dogiel$^3$
\\$^1$Department of Physics, The University of Hong Kong,
Pokfulam Road, Hong Kong, China
\\$^2$Institute of Astrophysics, Huazhong Normal
University, Wuhan 430079, China
\\$^3$I. E. Tamm Theoretical Physics
Division of P. N. Lebedev Institute, Leninskii pr, 53, 119991
Moscow, Russia} \maketitle

\begin{abstract}
The fluorescent X-ray emission from neutral iron in the molecular
clouds (Sgr B) indicates that the clouds are being irradiated by an
external X-ray source. The source is probably associated with the
Galactic central black hole (Sgr A*), which triggered a bright
outburst one hundred years ago. We suggest that such an outburst
could be due to a partial capture of a star by Sgr A*, during which
a jet was generated. By constraining the observed flux and the time
variability ($\sim$ 10 years) of the Sgr B's fluorescent emission,
we find that the shock produced by the interaction of the jet with
the dense interstellar medium represents a plausible candidate for
the  X-ray source emission.
\end{abstract}

\section{Introduction}

The major component in the Galactic center (GC) region is a central
molecular zone (Morris \& Serabyn 1996), which contains about 10\%
($\sim3\times 10^{7} M_{\odot}$) of the total molecular gas of the
Galaxy (Dahmen et al. 1998). The
most well-known and massive central molecular clouds are Sagittarius (Sgr) B1 and B2. The central
molecular zone is bright in diffuse X-rays (Sunyaev et al. 1993;
Koyama et al. 2007a), including thermal and non-thermal emission. The
thermal component, whose spectrum contains strong K-shell emission
lines from highly ionized atoms, such as Fe XXV (6.7 keV), and S XV
(2.45 keV), could be associated with a hot plasma in collisional
ionization equilibrium with multiple temperatures. This plasma
emission decreases monotonically with an increasing distance from
the GC, and extends about over 1 degree in longitude (Koyama et al.
1989; Yamauchi et al. 1990). In contrast, the brightest
non-thermal emission line from neutral iron, with a strong K-shell transition line at 6.4
keV,  was found to be near the Sgr B region (Koyama
et al. 1996). After the first discovery of the 6.4-keV line emission
by the {\it ASCA} satellite, a much deeper observation of the Sgr B
region was provided by the {\it Suzaku} satellite (Koyama et al.
2007b). Consequently, some remarkable features in the X-ray spectra
were observed, such as: a large equivalent width ($\geq$ 1~keV)
of the 6.4-keV line, and a deep iron K-absorption edge at 7.1~keV
(equivalent neutral hydrogen column density $\sim 10^{23-24}~\rm
cm^{-2}$).

These spectral features suggest that the 6.4-keV line, and the
underlying continuum emission, could originate from fluorescence and
Thomson scattering in the molecular clouds, which are irradiated by
an external X-ray source (Koyama et al. 1996; Sunyaev \& Churazov
1998). The detailed morphology of the 6.4-keV line emission of Sgr
B2 cloud was investigated by {\it Chandra} (Murakami et al.
2001). The line exhibits a concave shape (the peak position of the
emitting region is located about $1-2$ arcmin from the cloud center),
pointing to the GC direction. This strongly suggests that the
external source is loacted in the GC direction. However, no
sufficiently bright X-ray source has been detected there. Therefore,
Koyama et al. (1996, 2007b) and Murakami et al. (2001) proposed that
the supermassive black hole (BH), harbored at the GC (Sgr A*),
probably triggered a bright outburst about one hundred years ago,
which is the delayed time in the light travel due to the reflection
by Sgr B2. As an alternative to the above X-ray reflection model, Predehl
et al. (2003), Yusef-Zadeh et al. (2007), and Dogiel et al. (2009b)
proposed that the origin of the 6.4-keV line and X-ray continuum
emission could be due to non-relativistic electrons or protons.  However, this
 model is disfavored by the inferred metal
abundances of the hot plasma in the GC region (Nobukawa et al.
2010), even that this conclusion cannot be applied to the case of
protons, because their estimated abundance depends strongly on
the spectral index of the protons (see Dogiel et al. 2010). The most
convincing evidence that supports the X-ray reflection model comes
from the detection of a similar pattern of emission variability in
causally disconnected regions (Inui et al. 2009). This pattern was
recently confirmed by the observation of a clear decay, of about 40\% during the past 7
years, of the hard X-ray continuum emission from Sgr B2 (Terrier et
al. 2010). As reported by Ponti et al. (2010), such a time
variability can also be found in other molecular clouds (e.g., G
0.11-0.11). In particular, Ponti et al. (2010) observed an apparent superluminal
motion of a light front illuminating a molecular nebula, which
cannot be due to low energy cosmic rays, or to a source located inside
the cloud.

In this paper, we use the X-ray reflection model, and
suggest that the required past outburst could be due to a partial
capture of a star by Sgr A*. During the accretion of
the stellar matter onto the BH, an accompanying jet could be
generated, as seen in some accreting systems, such as microquasars
(e.g. Gallo et al. 2003; Fender 2003; Fender \& Maccarone 2004).
Then, by considering the deceleration of the jet by the dense interstellar
medium, the resulting shock emission could become a plausible X-ray
source, which is responsible for the fluorescent emission from the Sgr B
region. Such a (partial or full) stellar capture event by a central
supermassive BH has perhaps already been observed in some flaring
``normal" galaxies (e.g., Renzini et al. 1995; Li et al. 2002;
Donley et al. 2002). In our Galaxy, as suggested by Cheng et al.
(2006), stellar captures may also be required to provide a positron
source for the observed electron-positron annihilation emission from
the GC region. Meanwhile, in such a
stellar capture scenario, the diffuse hard X-ray and gamma-ray
emission from the GC region can also be explained well (Cheng et al. 2007; Dogiel et al.
2009a,b,c,d).

The present paper is organized as follows. In  Section \ref{2}, by using the X-ray
reflection model, we derive from the observed 6.4-keV line luminosity the X-ray luminosity of the external
source. In Section \ref{3}, we briefly describe the tidal capture
process.
% and qualitatively rule out the disk emission as the X-ray source.
In Section \ref{4}, we investigate the dynamics and the synchrotron
radiation of the shock produced by the deceleration of the jet.
Then, some basic properties of the jet are inferred from the
observations. Finally, we discuss and conclude our results in
Section \ref{5}.
%===============================================================================
%===============================================================================

\section{The X-ray reflection model}\label{2}
%\begin{table*}
%  \caption{The observational fluxes and luminosities of the 6.4-keV line emission from Sgr B2 cloud (M\,0.66$-$0.02)}
% \begin{tabular}{ccccccccc}
%      \hline\hline
%      Observatory & Suzaku & \multicolumn{2}{c}{XMM-Newton} & \multicolumn{2}{c}{XMM-Newton} & Chandra & ASCA \\
%      Detector & XIS(2005) & MOS(2004) & PN(2004) & MOS(2001) & PN(2001) & ACIS(2000) & SIS(1994) \\
%      \hline
%$F_{B,6.4}$ & $0.66_{-0.02}^{+0.02}$ & $0.58_{-0.04}^{+0.04}$ & $0.57_{-0.04}^{+0.04}$ & $0.92_{-0.11}^{+0.11}$ & $0.82_{-0.09}^{+0.10}$ & $1.00_{-0.06}^{+0.05}$ & $0.99_{-0.14}^{+0.13}$ \\
%$\mathcal C$    &      $1.05\pm 0.05$ & $0.91\pm 0.08$ & $0.89\pm 0.06$ & $1.25\pm 0.14$ & $0.83\pm 0.11$ & $0.99\pm 0.07$ & $1.19\pm 0.18$ \\
%$F_{B,6.4}^{\rm cor}$ & $0.63_{-0.04}^{+0.04}$ & $0.64_{-0.07}^{+0.07}$ & $0.64_{-0.06}^{+0.06}$ & $0.74_{-0.12}^{+0.12}$ & $0.99_{-0.17}^{+0.18}$ & $1.01_{-0.09}^{+0.09}$ & $0.83_{-0.17}^{+0.17}$ \\
%$\mathcal L_{B,6.4}$ &$0.45\pm 0.02$& $0.48\pm 0.05$& $0.48\pm 0.05$& $0.55\pm 0.09$& $0.73\pm 0.13$& $0.75\pm 0.07$& $0.62\pm 0.13$\\
%\hline
%\end{tabular}
%\\
%Note---The data of $F_{B,6.4}$ and $\mathcal C$ are taken from Inui
%et al. (2009). The observed line flux is in unit of $10^{-4}$
%photons cm$^{-2}$ s$^{-1}$ and the luminosity in unit of $10^{34}$
%erg s$^{-1}$.
%\end{table*}
In the following we specifically refer to Sgr B2 cloud as M $0.66-0.02$ (Inui et al. 2009), and a radius of $r_B\sim$3.2 arcmin is
adopted for this object.  The
distance to Sgr B2 is $d_B\approx7.8^{+0.8}_{-0.7}$ kpc (Reid et al. 2009), and consequently
$r_B\approx7.3$ pc.  Due to their projected offset $\sim
0.09$ kpc and to the offset along the light of sight, $\sim0.13$ kpc, the distance $d_{AB}$ between Sgr B2 and
Sgr A* is about $\sim0.16$ kpc (Reid et al. 2009). Denoting the (isotropically-equivalent) X-ray
luminosity of Sgr A* by $\mathcal L_{A,X}$, and the luminosity of the
6.4-keV line emission from Sgr B2 by $\mathcal L_{B,6.4}$, respectively, their ratio
can be roughly estimated as  (Murakami et al. 2000) %; Bambynek et al. 1972)
\begin{eqnarray}
\Re&\equiv&{\mathcal L_{B,\rm 6.4}\over \mathcal L_{A,X}}\nonumber\\
& =&{1\over4}\left({r_B\over d_{AB}}\right)^2e^{- \sigma(\geq7.1)
N_{\rm H,c} }\left(1-e^{-0.34\sigma_{\rm Fe}Z_{\rm Fe}N_{\rm
H,c}}\right)\nonumber\\
&&\times e^{-\sigma(6.4) N_{\rm
H,G}}\approx7.0\times10^{-6},\label{LAB relation}
\end{eqnarray}
where the reflection is assumed to be isotropic. In Eq.~(\ref{LAB relation}) $\sigma(E)$ is the
photoelectric absorption cross section per hydrogen atom for
standard interstellar matter
($\sigma(\geq7.1)\approx2.4\times10^{-24}\rm~cm^2$ and
$\sigma(6.4)\approx1.3\times10^{-24}\rm~cm^2$ (Morrison \& McCammon
1983)), $\sigma_{\rm Fe}\approx3.7\times10^{-20}\rm ~cm^{2}$ (for
$E\geq7.1$ keV) is the photoelectric absorption cross section of an
iron atom (Rakavy \& Ron 1967), the factor $0.34$ is the
fluorescence probability, and $Z_{\rm Fe}\approx1.6Z_{\rm
Fe,\odot}\approx10^{-5}$ is the abundance of iron (Nobukawa et al.
2010). The Galactic absorption from the cloud to the observer
is assumed to be $N_{\rm H,G}\approx6\times10^{22}\rm~H~cm^{-2}$
(Sakano et al. 1997), and from
the intensity of the CS emission, which is directly correlated with the
cloud mass, the column density of the Sgr B2 cloud can be
inferred to be $N_{\rm H,c}\approx8\times10^{23}~\rm H~cm^{-2}$  (Tsuboi et al. 1999; Ponti et al. 2010).
%(Murakami et al. 2000) or $\sim5\times10^{23}~\rm H~cm^{-2}$
%(Inui et al. 2009) from a fitting to the X-ray continuum spectra by
%a single power law with absorption,
The interstellar absorption from Sgr A* to the cloud is neglected. An improved model for the calculation of $\Re$ can be found in Murakami et al. (2000).

\begin{figure}
\resizebox{\hsize}{!}{\includegraphics{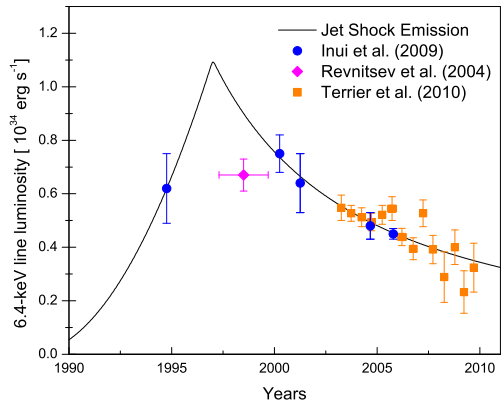}} \caption{The
observational luminosity data of the 6.4-keV line emission from Sgr
B2 cloud and an illustrative fitting (solid line) by using Equation
(\ref{nuLnu}). The fitting parameters are $t_0=1988$, $T_{\rm d}=9$
yr, and $\Re\mathcal L_{\rm
sh,X,d}=1.1\times10^{34}\rm~erg~s^{-1}$.}
\end{figure}

%The observed luminosity of the 6.4-keV line emission of Sgr B2 cloud
%can be calculated to be on the order of magnitude as
%\begin{eqnarray}
%\mathcal L_{B,6.4}=4\pi d_B^2\left(F_{B,6.4}\over \mathcal
%C\right)\sim10^{34}\rm~erg~s^{-1}, \label{LA}
%\end{eqnarray}
%where $F_{B,6.4}$ is the detected flux and $\mathcal C$ is a
%correction factor representing the detection efficiency for
%different observations.
There have been several deep exposure and short survey
observations of the Sgr B2 cloud with different satellites or
instruments, such as {\it ASCA, XMM-Newton, BeppoSAX, Chandra,
Suzaku}, and {\it INTEGRAL}. Some observational fluxes of the
6.4-keV line emission can be found from the literatures (e.g.,
Revnivtsev et al. (2004) and Inui et al. (2009)). They are of the order
of magnitude of $\sim10^{-4}~\rm photons~s^{-1}~cm^{-2}$. With a
distance $d_B\approx7.8^{+0.8}_{-0.7}$ kpc, the luminosity
$\mathcal L_{B,6.4}$ can be calculated to be around $\sim10^{34}~\rm
erg~s^{-1}$. The specific observational data are shown in Figure 1.
Combining this result with equation Eq.~(\ref{LAB relation}), we can estimate the order
of magnitude of the luminosity of the source as
\begin{eqnarray}
\mathcal L_{A,X}={\mathcal L_{B,6.4}\over \Re}\sim 10^{39}~\rm
erg~s^{-1},\label{lax}
\end{eqnarray}
which is much higher than the present luminosity of Sgr A*, which is of the order of
$\sim10^{33-35} \rm ~erg~ s^{-1}$ (Baganoff et al. 2003).
% as listed in Table 1. When we combine these
%different observations, the most serious problem is the reliability
%of the cross-calibration of their relative efficiencies. For this
%problem, Inui et al. (2009) suggested that the time-invariant
%6.67-keV line flux could be a good cross-calibration line, because
%the 6.67-keV line emission is due to the largely extended diffuse
%X-ray emission and a new supernova remanent G\,0.61+0.01. So the
%correction factors $\mathcal C$ can be defined as the ratios of the
%best-fit 6.67-keV line flux of each satellite to the averaged flux.
%Using this correction, we calculate the 6.4-keV line luminosities
%for different observations, which are listed in Table 1 and shown in
%Fig. 1 (solid circles).
Fig.~1 also shows some 6.4-keV line emission data, extrapolated
from the $20-60$ keV continuum emission data from Terrier et al.
(2010), by assuming that both the 6.4-keV line and the continuum
emissions have the same trend, and that the different sets of 6.4-keV
line emission data should be joined continuously. The apparent time
variation exhibited by the data indicates that the source emission
is likely to be a transient. This implies that about one hundred years ago a bright outburst
had happened at Sgr A*. The moment of the outburst can be obtained from the
delay in the light travel time, due to the reflection by Sgr B2, given by ${\rm \sim(0.16~ kpc-0.13~
kpc)}/c$ .

%===============================================================================
%===============================================================================

\section{Stellar tidal capture by Sgr A*}\label{3}

Observations show that at the center of Sgr A* there is a supermassive BH , with a
mass of $4\times10^6~M_{\odot}$ (see the review by Genzel \& Karas
2007). In the close vicinity of Sgr A* (less than 0.04 pc from it),
about 35 low-mass stars ($1-3~M_{\odot}$) and about 10 massive stars
($3-15~M_{\odot}$) are present (see Alexander \& Livio 2004). Hence, the
past outburst probably indicates a
(partial) capture of a star by Sgr A*. When a star gets very close
to a supermassive BH, it will experience tidal distortions,
and even disruptions. The strength of a tidal encounter can be
defined by the square root of the ratio between the surface gravity and
the tidal acceleration at the pericenter,
\begin{eqnarray}
\eta=\left({R_p^3\over GM_{\bullet}R_{\star}}{GM_{\star}\over
R_{\star}^2}\right)^{1/2},
\end{eqnarray}
where $G$, $R_p$, $M_{\star}$, $R_{\star}$, $M_{\bullet}$ are
Newton's constant, the pericentric distance, the stellar mass, the stellar
radius, and the BH's mass, respectively. If $M_{\bullet}>10^8~M_{\odot}$,
and for a sufficiently small $R_p$, the star would directly fall
into the BH's event horizon, without emitting any significant amount of energy. However, for a BH such as Sgr A*, the
star would disrupt when $\eta< 1$, which corresponds to
\begin{eqnarray}
R_{p}&<&R_T= R_{\star}\left({M_{\bullet}\over
M_{\star}}\right)^{1/3}\nonumber\\
&=&7.0\times10^{12}{~\rm cm}~\left({M_{\bullet}\over
10^6M_{\odot}}\right)^{1/3}\left({M_{\star}\over
M_{\odot}}\right)^{-1/3}\left({R_{\star}\over R_{\odot}}\right),
\end{eqnarray}
where $R_T$ is called the tidal radius. On the other hand, for somewhat
larger pericentric separations ($\eta>1$), the star avoids total
disruption, but it may lose parts of its envelope, which are captured by the
BH. The  stripped star, having positive energy, escapes from
the system. We denote the capture fraction of the total stellar mass
by $\xi$.

After the disruption or stripping, the bound material follows a
highly eccentric orbit, and after completing one orbit it returns to the central BH. A circular transient accretion disk forms, producing a transient black body emission. The emission reaches a peak
when most of the material first returns to the pericenter, after a time
$1.5~t_{\min}$, where (Ulmer 1999)
\begin{eqnarray}
t_{\min}&=&{2\pi R_p^3\over(GM_{\bullet})^{1/2}(2R_{\star})^{3/2}}\nonumber\\
&=&0.1~{\rm yr}~\left({R_p\over R_T}\right)^{3}\left({M_{\rm
BH}\over 10^6M_{\odot}}\right)^{1/2}\left({M_{\star}\over
M_{\odot}}\right)^{-1}\left({R_{\star}\over
R_{\odot}}\right)^{3/2}.\label{tmin}
\end{eqnarray}
%Such a timescale is obviously in disagreement with the time
%variability shown in Figure 1. Anyway,
We can use this timescale to estimate the accretion rate, and the
corresponding bolometric luminosity of the disk, as $\dot{M}=( 2\xi
M_{\star}/3t_{\min})(t/t_{\min})^{-5/3}$, and $\mathcal L_{\rm
disk}=\epsilon\dot{M}c^2$, respectively, where $\epsilon$ is the
radiation efficiency, which is usually taken to be $\sim$10\%.
However, for an accretion rate much lower than the Eddington rate,
the accretion flow is likely to be radiatively
inefficient (e.g., Yuan, Quataert \& Narayan 2003). In this case
most of the thermal energy released by viscosity, and increased by
compression, will be retained in the gas, and advected to the BH. Therefore
we could have a very small $\epsilon\ll$10\%. Moreover, since the
frequency with which a solar type star passes within a distance
$R_T$ is about $10^{-4}-10^{-5}\rm~ yr^{-1}$ (Rees 1988), the
capture event that happened one hundred years ago is quite unlikely to be
a full capture, i.e., $\xi\ll 1$.

Therefore the peak value of the disk luminosity can be estimated as (Ulmer 1999)
\begin{eqnarray}
\mathcal L_{\rm disk, p}&=&3.4\times10^{44}~{\rm erg~s^{-1}}~\left({\epsilon\over 0.01}\right)\left({\xi\over 0.1}\right)\left({R_p\over R_T}\right)^{-3}\nonumber\\
&&\times \left({M_{\bullet}\over
10^6M_{\odot}}\right)^{-1/2}\left({M_{\star}\over
M_{\odot}}\right)^{2}\left({R_{\star}\over R_{\odot}}\right)^{-3/2}.
\end{eqnarray}

Since this bolometric luminosity is very high, the disk emission
could be an appropriate source for the X-ray echo emission
of Sgr B2. Unfortunately, most of the disk
energy will be actually released into an energy band much lower than
hard X-rays, as can be seen in some stellar-capture-like flares, found in
UV/optical detections (e.g., Gezari et al. 2008, 2009). Moreover, a
UV flare detected at the center of a mildly active elliptical
galaxy, NGC 4552, even indicates a candidate for a partial capture
event (Renzini et al. 1995). Theoretically, we can also give an
upper limit for the disk temperature associated with an Eddington
luminosity, $\mathcal L_E=1.5\times10^{38}{\rm
erg~s^{-1}}(M_{\bullet}/M_{\odot})$, emitted from the tidal radius
(Ulmer 1999),
\begin{eqnarray}
T_{\rm eff}&=&\left({\mathcal L_E\over 4\pi
R_T^2\sigma}\right)^{1/4}\nonumber\\
&=&22~{\rm eV}~\left({M_{\bullet}\over
10^6M_{\odot}}\right)^{1/12}\left({M_{\star}\over
M_{\odot}}\right)^{1/6}\left({R_{\star}\over
R_{\odot}}\right)^{-1/2}
\end{eqnarray}
or from the Schwarzschild radius (Ulmer 1999),
\begin{eqnarray}
T_{\rm eff}=\left[{\mathcal L_E\over 4\pi
(5R_s)^2\sigma}\right]^{1/4}=48~{\rm eV}~\left({M_{\bullet}\over
10^6M_{\odot}}\right)^{-1/4},
\end{eqnarray}
where we have neglected a factor of the order of unity. The above equations may underestimate the real temperature in the disk, but the luminosity in the partial capture is also lower
than the Eddington luminosity. Therefore, it seems difficult for the
disk emission to provide sufficient X-ray photons with energy $>7.1$
keV to excite the strong 6.4-keV line emission of the clouds.

The strong UV emission can still influence the
environment of Sgr A*\footnote{We would like to point out in advance
that the UV emission could be generated about ten years earlier than
the hard X-ray emission, since the latter is probably produced by a
jet shock, whose emission reaches its peak about ten years after the
capture event (see the next section).}. Taking Sgr B2 as an example,
we can estimate the mean free path of a UV photon with energy
$\varepsilon_{\rm UV}\sim k_BT_{\rm eff}\sim 20$ eV in the cloud as
$l=(\sigma_a n_H)^{-1}= 2.3\times10^{13}$cm, where the number
density of neutral hydrogens is $n_H\sim
N_{H,c}/(2r_B)=1.8\times10^4~\rm cm^{-3}$, and the photoionization
absorption cross section can be expressed as
$\sigma_a\approx2.5\times10^{-18}(\varepsilon_{\rm UV}/20\rm
eV)^{-3}$ (Lang 1980). This indicates that if the layer
is so hot that it can be fully ionized, the UV photons can only
penetrate into a very thin surface layer of the cloud. In view of the small thermal
capacity of the layer, $\sim10^{41}\rm erg~K^{-1}$, it is indeed
very easy to heat the layer to high temperatures, as long as a small
fraction of the UV energy is converted into thermal energy. For example,
the ionized electrons can lose their kinetic energy to heat the
plasma. After being continuously irradiated by the UV flare for a few
months, the total mass of the fully ionized envelope of the cloud,
exposed to Sgr A*, can be estimated as $M_{\rm
ion}\sim{N_{UV}}\left({r_B^2/4d_{AB}^2}\right)m_H=25 M_{\odot}$,
where
$N_{UV}\sim (1.5L_{\rm disk, p}t_{\rm min}/\varepsilon_{\rm UV})$ is
the total number of the photons, $m_H$ is the mass of hydrogen
atom, and we assume that each UV photon ionizes one hydrogen atom. Consequently, the width of the ionized envelope is about
$\Delta\sim M_{\rm ion}/(2\pi r_B^2 n_H m_H)=5.2\times10^{14} {\rm
cm}= 23 ~l$, which means that only an extremely small fraction
$\Delta/r_B=2.3\times10^{-5}$ of the cloud can be ionized by the UV
flare. Finally, due to the high temperature of the ionized region,
the ionized material may be able to ultimately evaporate from the
cloud and mix with the surrounding plasmas.
%Actually, this could not be a
%serious problem, because the ionized clouds can reach recombination
%very quickly with a recombination timescale as $t_{\rm
%recom}=(\sigma_rn_ev_e)^{-1}< 2$ yr

%,where $\sigma_r>3\times10^{-6} v_e^{-2} \rm
%cm^2$ is the recombination cross section of a hydrogen-like ion
%(Lang 1980), , and $v_e\sim (k_BT/m_e)^{1/2}\approx 4\times10^6~\rm
%cm~s^{-1}$ is the typical electron velocity with $T\sim100$ K being
%the kinetic temperature of the clouds (de Vicente et al. 1997).}

What can be the appropriate X-ray source in this
stellar-partial-capture model? Similarly to the active galactic nuclei,
one may suggest that the Compton up-scattering of the soft disk
photons by a hot corona can produce strong hard X-ray emission. This
is a reasonable scenario, but unfortunately it is not clear
when and how a hot corona can be formed, and even whether there will
be a corona, since the disk is very short-lived and strongly
time-dependent. Even if a corona accompanying
the disk can indeed be formed, the temporal behavior of the corona emission still
should basically follow the behavior of the disk emission, whose
variability timescale (see Eq. 6) is however much shorter than 10
years. Alternatively, as suggested by Cheng et al. (2006), a jet
could be generated during the capture event. This can be suggested from
the fact that accreting BHs in microquasars sometimes are
accompanied by jet emission. Although in the high/soft state of
microquasars the jet formation is highly suppressed (Fender et al.
1999; Gallo et al. 2003), the evidences for jet emission are solid
enough in the low state (e.g. Gallo et al. 2003) and in the ``very high" state
(e.g. Fender 2003; Fender \& Maccarone 2004). Therefore, we suggest that the source emission
could be produced by the jet.

%===============================================================================
%===============================================================================

\section{Shock emission arising from the jet deceleration}\label{4}

If a jet was indeed ejected during the past capture event, a shock
would be produced from the interaction of the jet outflow with
the interstellar medium. Using such a jet shock emission, Wong et al.
(2007) explained the transient X-ray and optical data from some
nearby normal galactic nuclei, where a capture event may had also occurred.
%Therefore, in this section we attempt to (i)
%derive some basic properties of the shock from the observational
%data shown in Fig. 1 and (ii) find a physical process during which
%an appropriate shock could be produced.
%\subsection{The basic properties of the shock}
Due to the motion of the jet outflow, the ambient interstellar medium
is gradually swept up by the shock, and simultaneously the outflow is
decelerated. We denote the deceleration timescale of the outflow by
$T_{\rm d}$. Before $T_{\rm d}$, the outflow coasts, and due to the accumulation of the shocked
medium the emission of the shock increases. After $T_{\rm d}$, the emission is weakened due to
the significant deceleration of the outflow. In other words, the
shock emission would arrive at the peak at around $T_{\rm d}$, and
the increasing time of the emission roughly reflects the
deceleration timescale of the outflow (see Equation (\ref{nuLnu})).
Therefore, %if the transient X-ray emission of Sgr B region is indeed induced by a shock emission,
according to the observational data, shown in Fig. 1, %indicates that the peak time is between 1995 and 2000 and, in following estimations,
it seems acceptable to take decades as a fiducial value
for $T_{\rm d}$.

%The central molecular zone whose radius is about $100-200$ pc
%contains mass of $3\times10^7~M_{\odot}$. So we can roughly estimate
%the number density in this region to be $n\sim10^{2-3}~\rm cm^{-3}$.
Hence we can use $T_{\rm d}\sim10$ yr to find out whether initially the outflow
moves at a relativistic or non-relativistic velocity. For
an outflow with an isotropically-equivalent energy $\mathcal
E_{}$,\footnote{$\mathcal E_{}$ represents the total energy of the
outflow including its rest energy in the relativistic case, but only
the kinetic energy in the non-relativistic case.} we first define a
critical mass as $\mathcal M_{\rm of,c}=\mathcal
E/c^2=5.6\times10^{-5}M_{\odot}~\mathcal E_{ 50}$. Hereafter, the
convention $Q_x=Q/10^x$ is adopted in cgs units, except for the mass
that is expressed in units of $M_{\odot}$. Obviously, if the mass of the
outflow $\mathcal M_{\rm of}<\mathcal M_{\rm of,c}$, the outflow
 has initially a relativistic velocity, with a Lorentz factor of
$\Gamma_i=\mathcal E_{}/(\mathcal M_{\rm of}c^2)$. In this case, the
deceleration time can be defined as $\Gamma_i^2\mathcal M_{\rm
sw,d}c^2=\mathcal E_{\rm}$ with $\mathcal M_{\rm sw,d}={4\over 3}\pi
R_{\rm d}^3 nm_p$ and $R_{\rm d}=2\Gamma_i^2cT_{\rm
d}$,\footnote{The subscribe ``d" represents the values of the
quantities at the deceleration time $T_{\rm d}$. For a relativistic
shock, the internal energy of the shocked medium in its comoving
frame can be estimated by $E_{\rm in}=(\Gamma_i-1)\mathcal M_{\rm
sw,d}c^2$ according to the shock jump condition (Blandford \& Mckee
1976). So the total energy of the shocked medium can be expressed by
$\mathcal E=\Gamma_i(E_{\rm in}+M_{\rm sw,d}c^2)=\Gamma_i^2\mathcal
M_{\rm sw,d}c^2$.} where $n$ is the number density of the ambient
medium, $\mathcal M_{\rm sw}$ is the total mass of the swept-up medium,
$R$ is the radius of the shock, $m_p$ is the mass of proton, and $c$ is the
speed of light. The above equations yield
\begin{eqnarray}
T_{\rm d}=\left({3\mathcal E_{}\over32\pi
nm_pc^5\Gamma_i^8}\right)^{1/3}= 2.0\times10^4{\rm~ s}~\mathcal E_{
50}^{-7/3}\mathcal M_{\rm of,-5}^{8/3}n_{1}^{-1/3},
\end{eqnarray}
where we take $n\sim10~\rm cm^{-3}$ as a fiducial
value\footnote{Actually the gas density distribution in the GC
region is complicated. According to Jean et al. (2006), the bulge
region inside the radius $\sim230$ pc and height 45 pc contains
$7\times10^7~M_{\odot}$. A total of 90\% of this mass is trapped in
small high density clouds (as high as $10^{3}~\rm cm^{-3}$), while
the remaining 10\% is homogeneously distributed with an average
density $\sim10~\rm cm^{-3}$.}. On the other hand, if $\mathcal
M_{\rm of}>\mathcal M_{\rm of,c}$, the outflow would be
non-relativistic, with a velocity of $V_i=(2\mathcal E_{}/\mathcal
M_{\rm of})^{1/2}$. In this case, we can obtain $T_{\rm d}$ from
the equations ${1\over2}\mathcal M_{\rm sw,d}V_i^2=\mathcal
E_{},~\mathcal M_{\rm sw,d}={4\over 3}\pi R_{\rm d}^3 nm_p$, and
$R_{\rm d}=V_iT_{\rm d}$, respectively, as
\begin{eqnarray}
T_{\rm d}=\left({3\mathcal E_{}\over2\pi nm_pV_i^5}\right)^{1/3}=
1.4\times10^9{\rm~ s}~\mathcal E_{ 50}^{-1/2}\mathcal M_{\rm
of,-1}^{5/6}n_{1}^{-1/3}.\label{tdec2}
\end{eqnarray}
For $T_{\rm d}\sim10{~\rm yr}$, the relativistic case requires
\begin{eqnarray}
\mathcal E_{}>4.3\times10^{56}{\rm~ erg}~n_1T_{\rm d,8.5}^{3},
\end{eqnarray}
a condition which is physically unacceptable. In contrast, the non-relativistic
case yields a more reasonable result,
\begin{eqnarray}
V_i(\mathcal E_{})&=&2.5\times10^{9}~{\rm cm~s^{-1}}~\mathcal
E_{50}^{1/5}n_1^{-1/5}T_{\rm
d,8.5}^{-3/5},\label{vejh}\\
\mathcal M_{\rm of}(\mathcal
E_{})&=&1.7\times10^{-2}~M_{\odot}~\mathcal E_{
50}^{3/5}n_1^{2/5}T_{\rm d,8.5}^{6/5},\label{mejh}
\end{eqnarray}
which is only mildly dependent on the uncertain model
parameter $\mathcal E_{}$.
% In the 500 pc region the average
%density will drop to $\sim(1-3)~\rm cm^{-3}$.
%===============================================================================
%===============================================================================

%\subsection{Synchrotron X-ray radiation from the shock}
The dynamic evolution of the non-relativistic adiabatic shock is determined by
the energy conservation law ${1\over2}\mathcal M_{\rm
sw}V^2=\mathcal E$, which gives
\begin{equation}
V(t)=V_i\left\{
\begin{array}{ll}
\tilde{t}^0,&\tilde{t}<1, \\
\tilde{t}^{-3/5},&\tilde{t}>1
\end{array}\right.,
\end{equation}
where $\tilde{t}\equiv (t-t_0)/T_{\rm d}$, with $t_0$ being the
initial time of the capture event. The deceleration timescale
$T_{\rm d}$ is defined by Eq.~(\ref{tdec2}). This is the
well-known Sedov solution (Sedov 1969). When the shock propagates
through the medium, the energy density behind the shock is given by (Lang
1980)
\begin{eqnarray}
e={9\over8}nm_pV^2.
\end{eqnarray}
By assuming that a fraction  $\epsilon_B$ of the internal energy is in form of magnetic energy, the magnetic field strength can be
estimated as $B=(8\pi \epsilon_Be)^{1/2}$. The shock-accelerated
electrons are assumed to have a power-law distribution of the kinetic
energy $(\gamma-1)m_ec^2$, with a minimum Lorentz factor $\gamma_m$ given by
(Huang \& Cheng 2003)
\begin{eqnarray}
{dN_e\over d\bar{\gamma}} \propto {\bar{\gamma}}^{-p},&~{\rm for~}
\bar{\gamma}\geq\bar{\gamma}_m,
\end{eqnarray}
where for simplicity $\bar{\gamma}\equiv\gamma-1$ is defined as an effective
Lorentz factor,  and $p\simeq2.2$  is the spectral index (Gallant et al.
2002). If the internal energy of the electrons is a small fraction $\epsilon_e\sim 0.1$ of the total internal
energy (smaller than that of the  protons), we can obtain the value of $\bar{\gamma}_m$ as (Dai \& Lu
2001)
\begin{equation}
\bar{\gamma}_m={9\over32}\epsilon_e{p-2\over p-1}{m_p\over
m_e}\left(V\over c\right)^2 =\bar{\gamma}_{m,\rm d}\left\{
\begin{array}{ll}
\tilde{t}^0,&\tilde{t}<1, \\
\tilde{t}^{-6/5},&\tilde{t}>1
\end{array}\right.,\label{gm}
\end{equation}
where $\bar{\gamma}_{m,\rm d}=0.06~\epsilon_{e,-1}\mathcal E_{
50}^{2/5}n_{1}^{-2/5}T_{\rm d,8.5}^{-6/5}$. Considering the
synchrotron cooling\footnote{Following Sari \& Esin (2001), we can
estimate the radiation density as
$u_{\gamma}=(\gamma_c/\gamma_m)^{2-p}\epsilon_ee$ (representing here an upper limit
), while the energy density of the magnetic fields is
$u_{B}=B^2/8\pi=\epsilon_Be$.
%=0.02{\rm~G}~\epsilon_{B,-1}^{1/2}\mathcal E_{50}^{1/5}n_1^{3/10}T_{\rmd,8.5}^{-3/5}
Then the ratio of the luminosity of the inverse-Compton radiation to the
synchrotron luminosity is given by
$u_{\gamma}/u_B=(\epsilon_e/\epsilon_B)(\gamma_c/\gamma_m)^{2-p}\sim0.1$,
which is much smaller than one. Therefore, in this paper
we would not consider the inverse-Compton scattering of the
electrons.} of the electrons with power
$P(\gamma)={4\over3}\sigma_Tc(\gamma^2-1){B^2\over8\pi}$, a critical
cooling Lorentz factor can be determined by $\bar{\gamma} m_e
c^2=P(\gamma) t$ to be
\begin{equation}
\gamma_c={6\pi m_e c\over\sigma_TB^2t}=\gamma_{c,\rm d}\left\{
\begin{array}{ll}
\tilde{t}^{-1},&\tilde{t}<1, \\
\tilde{t}^{1/5},&\tilde{t}>1
\end{array}\right.,\label{gc}
\end{equation}
where $\gamma_{c,\rm d}=8.5\times10^3~\epsilon_{B,-1}^{-1}\mathcal
E_{ 50}^{-2/5}n_{1}^{-3/5}T_{\rm d, 8.5}^{1/5}\gg1$. Within the time $t$ an electron
with an initial Lorentz factor $\gamma>\gamma_c$ would quickly cool
down  below $\gamma_c$. Therefore, the
spectral index of the net electron distribution above $\gamma_c$
becomes $p+1$, and a radiation spectrum of the form $\nu^{-p/2}$ can be
produced by these radiative electrons (Sari et al. 1998).

The peak spectral power of an electron can be obtained from the ratio of
the total power and of the synchrotron characteristic frequency
(Sari et al. 1998)
\begin{equation}
P_{\nu,
\max}\approx{P(\gamma)\over\nu(\gamma)}={\sigma_Tm_ec^2\over3q}B\beta^2.\label{pnu}
\end{equation}
$P_{\nu,
\max}$ is independent of $\gamma$. Then, for $\nu>\nu_c$,  the normalized radiation
spectrum can be written as
\begin{eqnarray}
\mathcal L_{\rm sh}=\nu
{N_{e,>\gamma_c}P_{\nu,\max}}\left({\nu\over\nu_c}\right)^{-p/2},\label{Lshspectrum}
%&=&\nu \left[{N_{e,\rm tot}}\left({\bar{\gamma}_c\over\bar{\gamma}_m}\right)^{-(p-1)}\right]P_{\nu,\max}\left({\nu\over\nu_c}\right)^{-p/2}
\end{eqnarray}
where the characteristic frequency, corresponding to the
$\gamma_c$-electrons, is given by
\begin{equation}
\nu_c={q\over2\pi m_ec}B\gamma_c^2=\nu_{c,\rm d}\left\{
\begin{array}{ll}
\tilde{t}^{-2},&\tilde{t}<1, \\
\tilde{t}^{-1/5},&\tilde{t}>1.
\end{array}\right.\label{nuc}
\end{equation}
where $\nu_{c,\rm d}=3.5\times10^{12}~{\rm
Hz}~\epsilon_{B,-1}^{-3/2}\mathcal E_{ 50}^{-3/5}n_1^{-9/10}T_{\rm
d, 8.5}^{-1/5}\ll \nu_X\sim10^{18}$ Hz. The spectral index $p/2$ is
perfectly consistent with the observational index (Terrier et al.
2010). The number of the electrons above $\gamma_c$ can be obtained as
\begin{eqnarray}
N_{e,>\gamma_c}={4\over3}\pi
R^3n\left({\bar{\gamma}_c\over\bar{\gamma}_m}\right)^{-(p-1)}.
\end{eqnarray}
For a fixed frequency $\nu_X\sim 10^{18}$ Hz, Eq.~
(\ref{Lshspectrum}) gives the evolution of the X-ray luminosity as
\begin{equation}
\mathcal L_{\rm sh,X}=\mathcal L_{\rm sh,X,d}\times\left\{
\begin{array}{ll}
\tilde{t}^{2},&\tilde{t}<1, \\
\tilde{t}^{-(3p-4)/2},&\tilde{t}>1
\end{array}\right.,\label{nuLnu}
\end{equation}
where the peak value at $T_{\rm d}$ is
\begin{eqnarray}
\mathcal L_{\rm sh,X,d}&=&7.8\times10^{37}{~\rm
erg~s^{-1}}~\epsilon_{e,-1}^{p-1}\epsilon_{B,-1}^{(p-2)/4}\nonumber\\
&&\times \mathcal E_{
50}^{p/2}n_{1}^{(2-p)/4}\nu_{X,18}^{(2-p)/2}T_{\rm d,
8.5}^{(4-3p)/2}.\label{lxdec}
\end{eqnarray}
 The  fitting to the observational data, obtained by using Eq.~(\ref{nuLnu}), is shown by the solid line in Fig. 1, where
$\mathcal L_{\rm sh,X,d}=(1.1\times10^{34}/\Re) ~\rm erg~s^{-1}$.
The confrontation between the model and the observations shows that
most of the data can be well explained by this simple jet shock
emission model. Only the {\it BeppoSAX} data (diamond)
do not fit with the fitting line.

Taking into account that the jet may be concentrated within a narrow
cone, we introduce a beaming factor $f_b$, and denote the
beaming-corrected energy of the jet by $E_{\rm jet}$. Then we can
solve the equation $f_b\mathcal L_{\rm sh,X,d}(\mathcal E)=\mathcal
L_{A,X,\rm peak}\sim 10^{39}~\rm erg~s^{-1}$ with $\mathcal E=E_{\rm
jet}/f_b$ to obtain
\begin{eqnarray} E_{\rm
jet}&=&0.8\times10^{51}{\rm erg}~f_{b,-1}^{(p-2)/p}
\epsilon_{e,-1}^{2(1-p)/p}\epsilon_{B,-1}^{(2-p)/2p}\nonumber\\
&& \times\mathcal L_{A,X,\rm
peak,39}^{2/p}n_1^{(p-2)/2p}\nu_{18}^{(p-2)/p}T_{\rm
d,8.5}^{(3p-4)/p},
\end{eqnarray}
where the emission from the beamed jet is assumed to be isotropic.
Moreover, substituting $\mathcal E=E_{\rm jet}/f_b$ into Eqs.~
(\ref{vejh}) and (\ref{mejh}), we further obtain
\begin{eqnarray}
V_{i}& =&6.0\times10^{9}{\rm
cm~s^{-1}}~f_{b,-1}^{-2/5p},\\%\epsilon_{e,-1}^{2(1-p)/5p}\epsilon_{B,-1}^{(2-p)/10p},\\%~,\\%\mathcal L_{A,X,38}^{2/5p}T_{\rm d,8.5}^{(3p-4)/5p}n_1^{(p-2)/10p},\\
M_{\rm
jet}&=&f_{b}\mathcal M_{\rm of}=0.02~M_{\odot}~f_{b,-1}^{(5p-6)/5p},%\epsilon_{e,-1}^{6(1-p)/5p}\epsilon_{B,-1}^{3(2-p)/10p}.%.%\mathcal L_{A,X,38}^{6/5p}T_{\rm d,8.5}^{(9p-6)/5p}n_1^{3(p-2)/10p}.
\end{eqnarray}
where $\epsilon_{e,-1}=\epsilon_{B,-1}=\mathcal L_{A,X,\rm
peak,39}=n_1=\nu_{X,18}=T_{\rm d,8.5}=1$.
%The uncertainty of these jet parameters are mainly from the unknown
%beaming factor.
By assuming that a fraction $\zeta$ (typically $1-10$\%; Yuan et al.
2005) of the total accreted stellar matter can be ejected into the
jet, the mass loss of the star can be estimated to be
$\sim0.2~M_{\odot}~f_{b,-1}^{(5p-6)/5p}\zeta_{-1}^{-1}$, which can
be reduced if the jet emission is anisotropic. All of the above
results show that one can choose the jet as the hard X-ray source.

It should be noticed that the above analytical calculations are some
rough approximations. For example, the realistic transition in the
light curve from the increasing phase to the decreasing phase would
be much smoother. In this case the {\it BeppoSAX} data could be
understood. Moreover, we do not consider the sideway expansion of
the jet. If we take ${c_s={1\over4} V_i}$ as the upper limit of the
sound speed of the shocked medium, the timescale of the sideway
expansion of the jet can be estimated as $2\sqrt{f_b}R_d/c_s\sim
2T_d$. This timescale indicates that about few decades later the jet
will become nearly isotropic diffuse material, which will appear as a
new component of the environment of Sgr A*, within a region of a few
tens of arcseconds. Hence, it is nearly impossible to image the jet one
hundred years after the capture event. Nevertheless, on much longer
timescales, the interaction of these BH-ejected diffuse materials
with the initial diffuse surrounding medium could still play an
important role in the GC diffuse X-ray, gamma-ray, 511 keV
annihilation line emissions, and in the heating of the surrounding
plasma, which had been  thoroughly studied in Cheng et al. (2006,
2007) and Dogiel et al. (2009a, b, c).

In addition to Sgr B2, the jet emission can, at least in principle, also
influence the other molecular clouds around Sgr A*, but such an
influence mostly depends on the distances between the clouds and
Sgr A*. Consequently, some clouds reflect the increasing emission,
while some others reflect the decreasing emission. More importantly,
the emission from the beamed jet is probably highly anisotropic, at
least during the first decades, although for simplicity the isotropic approximation
is adopted in our calculations. So the reflection by
different clouds could be strongly dependent on their viewing angles
with respect to the direction of motion of the jet. In addition,
the different properties and environments of the different clouds,
can also lead to a great diversity of the reflection
emission. To a certain extent, Sgr B2 could have a very
particular position, just in front of the jet. The
distance of the cloud to Sgr A* makes it possible to detect both the
increasing and the decreasing emission phases.

%\begin{table*}
%\caption{The inferred properties of the outflow}
%\begin{tabular}{cccc}
%\hline\hline
%$f_b$ & $E/10^{50}$ erg  &$M_{\rm of}/M_{\odot}$&$V_i/c$\\
%\hline
%1     & $0.35-2.8$ & $0.010-0.036$ & $0.065-0.098$   \\
%0.1   & $0.28-2.3$ & $0.036-0.125$ & $0.098-0.149$  \\
%0.01  & $0.23-1.9$ & $0.125-0.438$ & $0.149-0.227$ \\
% \hline
%\end{tabular}
%\end{table*}
%===============================================================================
%===============================================================================

\section{Conclusion and discussion}\label{5}

In the framework of the X-ray reflection model for the 6.4-keV line
emission from Sgr B, we propose that the external X-ray source could
be associated with a stellar partial capture event at Sgr A*. A
qualitative comparison between the observational and theoretical
timescales and luminosities further shows that the source emission
is likely to be produced by the shock produced by the jet
deceleration (but not by the accretion disk). The inferred energy,
mass, and velocity of the jet show that (i) the jet has a
low-energy ($\sim10^{51}\rm erg$), and it is non-relativistic ($\sim0.1c$)
and (ii) the star could be only partially stripped, rather than
totally disrupted, corresponding to a capture fraction of $\xi\sim
0.2~M_{\star,0}^{-1}f_{b,-1}^{(5p-6)/5p}\zeta_{-1}^{-1}$.

It has been suggested for a long time that supermassive BHs in relatively low
luminosity active galactic nuclei can be fed by the tidal capture of
stars. However, direct observational evidence for such capture
processes is lacking. The investigation in this paper shows
that, in view of the short distance to the GC, it may be worthwhile
and feasible to carefully observe the GC region to find clues to some
historical capture events, even though now Sgr A* is quiescent, with an
X-ray luminosity of only $\sim10^{33-34}\rm~erg~s^{-1}$ (Baganoff et
al. 2003).

Finally, we would like to point out some alternative
scenarios for the X-ray source, e.g., (i) Fryer et al. (2006)
suggested that the X-ray source could be due to a supernova shock
hitting the $50 \rm km~ s^{-1}$ molecular cloud behind, and to the
east, of Sgr A*. The last vestige of this interaction is visible now
as Sgr A East; (ii) Cuadra et al. (2008) found that shocks produced
by stellar winds can create cold clumps of gas, the accretion of
which onto Sgr A* would produce for a decade intervals of activity with luminosity
as high as $10^{39} ~\rm erg~ s^{-1}$. The wind-clump-capture model may have some similarities to
our stellar-partial-capture model. But in the present paper we have considered a more detailed analysis of the physical processes after the matter capture by the black hole, and which could be directly responsible for
the X-ray emission.

\section*{Acknowledgements}

We thank Dr. Tong Liu for useful discussions, Dr. T. Harko for a
critical reading of the manuscript, and the anonymous referee for
valuable suggestions that helped us to improve the paper. This work
is supported by the GRF Grants of the Government of the Hong Kong
SAR under HKU 7011/10P. YWY is also partly supported by the National
Natural Science Foundation of China (grant 10773004). DOC and VAD
are supported by the RFBR grant 08-02-00170-a, the NSC-RFBR Joint
Research Project RP09N04 and 09-02-92000-HHC-a.

\end{document}